\documentstyle[epsfig]{aipproc}
\def\lla{\left\langle}
\def\rra{\right\rangle}

\begin{document}

%% net cover
%\addtolength{\baselineskip}{.2cm}

\begin{flushright}
UR-1531\\
ER/40685/918\\
VPI-IPPAP-98-2\\
May 1998
\end{flushright}

\vspace*{.2in}

\begin{center}
{\bf Two-Higgs-Doublet-Models and Radiative CP Violation}$^\star$

\vspace*{.4in}
{\bf Otto C.W. Kong$^*$ and Feng-Li Lin$^{\dagger}$ }

\vspace*{.4in}
{\it $^*$Department of Physics and Astronomy,\\
University of Rochester, Rochester NY 14627-0171.\\[.2in]
$^{\dagger}$ Department of Physics, and Institute for Particle Physics and Astrophysics,\\ 
Virginia Polytechnic Institute and State University, Blacksburg, VA 24061-0435.}

\vspace*{.8in}
{Abstract}
\end{center}

\noindent
We discuss the feasibility of spontaneous CP violation being
induced by radiative corrections in 2HDM's. Specifically, we analyze the
cases of gaugino/higgsino effect on MSSM, and a new model with an extra
exotic quark doublet. The new model, while demonstrating well the
Georgi-Pais theorem, is also expected to be phenomenlogically interesting.

\vfill
\noindent --------------- \\
$^\star$ Talk given by O.K. at MRST conference, McGill University,
Montr\'eal, May 1998 --- submission for proceedings.  

\clearpage

%%%%%%%%%%%%%%%%%%%%%%%%%%%%%%%%%%%%

%\addtolength{\baselineskip}{-.2cm}
\addtocounter{page}{-1}

\title{Two-Higgs-Doublet-Models\\ and Radiative CP Violation
\thanks{Talk presented by O.K..}}

\author{Otto C.W. Kong$^*$ and Feng-Li Lin$^{\dagger}$}
\address{$^*$Department of Physics and Astronomy,\\
University of Rochester, Rochester NY 14627-0171.\\[.2in]
$^{\dagger}$ Department of Physics, and Institute for Particle Physics and Astrophysics,\\ 
Virginia Polytechnic Institute and State University, Blacksburg, VA 24061-0435.}

%\lefthead{LEFT head}
%\righthead{RIGHT head}
\maketitle

\begin{abstract}
We discuss the feasibility of spontaneous CP violation being
induced by radiative corrections in 2HDM's. Specifically, we analyze the
cases of gaugino/higgsino effect on MSSM, and a new model with an extra
exotic quark doublet. The new model, while demonstrating well the
Georgi-Pais theorem, is also expected to be phenomenlogically interesting.  
\end{abstract}

\section*{Introduction}
The source of CP violation is one of the most important unsolved puzzles 
in particle physics. On the one hand, CP violation is  observed
experimentally only in the $K^0$-$\bar{K}^0$ system, with the corresponding 
weak CP phase compatible with the Kobayashi-Maskawa (KM) mechanism.
On the other hand, the experimental bound on the neutron electric dipole
moment indicates that the effective strong CP phase has to be 
exceedingly small,
\begin{equation}
\bar{\theta} <   10^{-9}~.
\end{equation}
The origin of this strong CP problem
lies in the necessity of adding the so-called $\theta$ term to the
effective QCD Lagrangian due to the contribution of instantons
present in the topologically nontrivial QCD vacuum :
\begin{equation}
{\mathcal L}_{{\mathrm eff}} = \frac{\theta\alpha_s}{8\pi}
F^A_{\mu\nu}\tilde{F}^{A\mu\nu}~,
\end{equation}
where the dual field strength is given by $\tilde{F}_{\mu\nu} = \frac{1}{2}
\epsilon_{\mu\nu\alpha\beta}F^{\alpha\beta}$.
Through the anomaly in the QCD axial ${\mathrm U}(1)$ current,
chiral U(1) transformations lead to shifts in $\theta$, leaving
the physical combination $\bar{\theta} = \theta - {\mathrm arg\, det} M_q$,
where $M_q$ is the quark mass matrix, as the effective strong CP phase.
In the supersymmetrized version of the standard model, new CP phases
from supersymmetry (SUSY) breaking terms also have to be small
($< 10^{-3}$). Spontaneously CP violation (SCPV) provides an elegant
theory with a good control on the magnitude of the various CP phases.
Hence models of SCPV keep generating new interest.

\section*{SCPV in a two-Higgs-doublet model }
The most simple setting for achieving the SCPV scenario is given by a
two-Higgs-doublet model (2HDM)\cite{lee}.
The most general scalar potential for two Higgs doublets,
$\phi_1$ and $\phi_2$, is given by
\begin{eqnarray}
V(\phi_1, \phi_2) &=& m_1^2 \phi_1^{\dag} \phi_1 + m_2^2 \phi_2^{\dag} \phi_2 - (m_3^2 \phi_1^{\dag} \phi_2 + h.c.) 
\nonumber\\
&&+ \lambda_1 (\phi_1^{\dag} \phi_1)^2 + \lambda_2 (\phi_2^{\dag} \phi_2)^2 + \lambda_3 (\phi_1^{\dag} \phi_1)(\phi_2^{\dag} \phi_2)
\nonumber\\
&&+ \lambda_4 (\phi_1^{\dag} \phi_2)(\phi_2^{\dag} \phi_1) + \frac{1}{2} \lbrack \lambda_5 (\phi_1^{\dag} \phi_2)^2 + h.c. \rbrack
\nonumber\\
&&+ \frac{1}{2} \lbrace \phi_1^{\dag} \phi_2 \lbrack \lambda_6 (\phi_1^{\dag} \phi_1) + \lambda_7 (\phi_2^{\dag} \phi_2) \rbrack + h.c. \rbrace~.
\end{eqnarray}
Assuming  all the parameters in $V$ being real, and denoting the
vacuum expectation values (VEV's) of the neutral components of the Higgs doublets   by
\[ 
\langle \phi_1^0 \rangle = v_1 \qquad {\rm and}  \qquad
\langle \phi_2^0 \rangle = v_2 e^{i\delta}~,
\]
we have 
\begin{eqnarray}
\left\langle V\right\rangle &=& m_1^2 v_1^2 + m_2^2 v_2^2 + \lambda_1 v_1^4 + \lambda_2 v_2^4 + (\lambda_3 + \lambda_4 - \lambda_5) v_1^2 v_2^2 
\nonumber\\
&&+ 2 \lambda_5 v_1^2 v_2^2 \cos^2\delta -(2 m_3^2 - \lambda_6 v_1^2 -\lambda_7 v_2^2) v_1 v_2 \cos\delta~,
\nonumber\\
&=& M_1 v_1^2 + M_2 v_2^2 + (p v_1^4 + 2r v_1^2 v_2^2 +q v_2^4)
\nonumber\\
&&+ 2 \lambda_5 v_1^2 v_2^2 (\cos\delta -\Omega)^2-\frac{m_3^4}{2 \lambda_5}~;
\end{eqnarray}
where
\begin{equation}
\Omega = \frac{2m_3^2 - \lambda_6 v_1^2 - \lambda_7 v_2^2}{4 \lambda_5 v_1 v_2}~,
\end{equation}
and
\begin{eqnarray}
M_1 &=& m_1^2 + \frac{\lambda_6 m_3^2}{2 \lambda_5}~,
\\
M_2 &=& m_2^2 + \frac{\lambda_7 m_3^2}{2 \lambda_5}~,
\\
p &=& \lambda_1 - \frac{\lambda_6^2}{8 \lambda_5}~,
\\
q &=& \lambda_2 - \frac{\lambda_7^2}{8 \lambda_5}~,
\\
r &=& \frac{1}{2}(\lambda_3 + \lambda_4 - \lambda_5 - \frac{\lambda_6 \lambda_7}{4 \lambda_5})~.
\end{eqnarray}
A nontrivial phase ($\delta$) then indicates SCPV. 

Let us look at the $\delta$-dependence of $\left\langle V\right\rangle$. 
The extremal condition gives
\begin{equation}
 -4 \lambda_5 v_1^2 v_2^2(\cos\delta - \Omega)sin\delta =0 \; ,
\end{equation}
and the stability condition requires
\begin{equation}
\frac{\partial^2 V}{\partial \delta^2}=4\lambda_5 v_1^2 v_2^2 \lbrack \cos\delta(\Omega-\cos\delta ) + sin^2\delta \rbrack > 0 \; .
\end{equation}
$\cos\delta = \Omega$ gives a SCPV solution, provided that 
$\lambda_5 > 0$ and $|\Omega|< 1$.
Actually,  Eq.(2) shows that this is the absolute minimum.
In order for V to have a lower bound, we have the extra constraints
\begin{equation}
p > 0~,\;\;\;\;\;\;\; q > 0~,
\end{equation}
and
\begin{equation}
pq >  [r+\lambda_5 (\cos\delta -\Omega)^2]^2~.
\end{equation}
The latter reduced to 
\begin{equation}
pq >  r^2
\end{equation}
for the CP violating minimum.

However, in order to 
avoid flavor-changing-neutral-currents that could result, extra structure
like natural flavor conservation (NFC)\cite{nfc} has to be imposed on
a 2HDM, which then forbids  SCPV\cite{brc}. For instance, a natural way to impose NFC is to require that only one of the Higgs, say $\phi_1$, 
transforms nontrivially under an extra discrete
symmetry. This means that $m_3^2$, $\lambda_6$ and $\lambda_7$,
and may be  $\lambda_5$ too, all have to vanish. Similarly,
a supersymmetric version of the standard model (SM) is naturally a 2HDM
with NFC being imposed automatically by the holomorphy 
of the superpotential. The tree level scalar potential there has vanishing
$\lambda_5$, $\lambda_6$, and $\lambda_7$, though the soft SUSY breaking
$B$-term gives rise to a nonvanishing  $m_3^2$.  The interesting point of 
concern then is whether radiative corrections can modify  the  picture.
Note that a positive $\Delta\lambda_5$ is needed for this radiative CP
violation scenario.

\section*{Radiative CP violation} 
\vspace*{.95in}

%\twocolumn
\begin{figure}[b]
%\vspace{10in}
\includegraphics{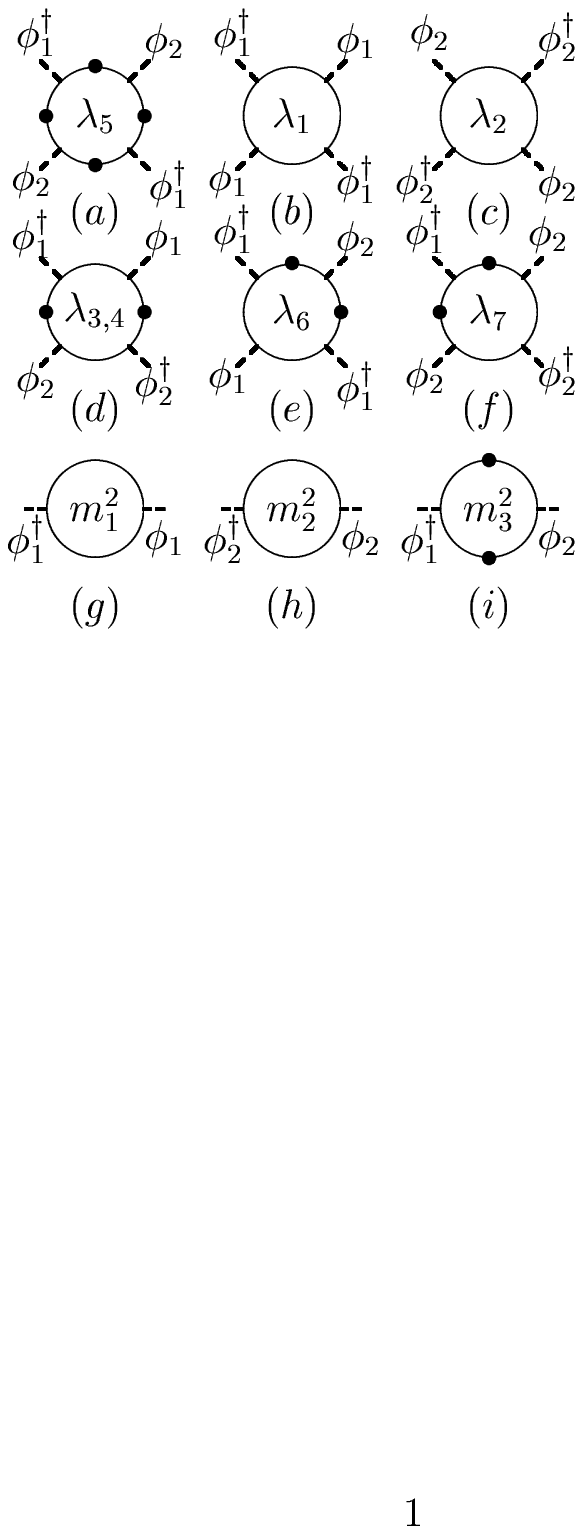}
%\vspace{1in}
\end{figure}

\vspace*{-1in}
For the case of the minimal supersymmetric standard model (MSSM), 
Maekawa\cite{mae} pointed out that there
is a positive contributions to $\lambda_5$ from a finite 1-loop diagram 
(Fig. 1a) involving the gauginos and higgsinos, which could lead to 
SCPV. However, there are objections to the particular scenario. Maekawa\cite{mae}
realized that the situation will not be able to give rise to sufficient weak
CP vioation for the $K^0$-$\bar{K}^0$ system. Pomarol\cite{pol} 
argued that  mass for the $``$psuedoscalar", $m_A$,
is roughly proportional to $\sqrt{\lambda_5}$ and is at least
more than a factor of three too small to be phenomenologically
acceptable. Actually, we worked out the algebra to decoupled the Goldstone
mode explicitly assuming the CP violating vacuum, and got the following
results\cite{paper}: the $3\times 3$
physical Higgs mass-squared matrix $m^2_{Hij}$ gives
\begin{equation}
\det{(m_H^2)}=\lambda_5 (pq-r^2)sin^2 2\beta sin^2\delta \; ;
\end{equation}
\hfill \rule{3in}{.02in} \hspace*{.3in}\\
\begin{minipage}{2in}
as $p=q=-r>0$ at tree level, when only a positive $\Delta\lambda_5$ is 
considered, $(pq-r^2)$ and hence $\det{(m_H^2)}$ becomes
negative [see Eq.(8)].  In fact, with a negative $r$, $(pq<r^2)$ exactly 
violates the condition for  the potential to be bounded from below.
This inconsistency is also pointed out by Haba\cite{hb}, who then 
suggested that  it would be fixed when top/stop loop contributions to the 
other parameters, mainly $\lambda_2$, in the potential are included.
From our perspective, in order to
have a {\it consistent approximation} to the radiative corrections in
whatever interesting region of the parameter space, loop contributions
to {\it all} parameters in the scalar potential at the same order have 
to be considered. Corrections to all parameters in $V$ {\it do} exists, 
as shown in Fig. 1.
\end{minipage}

\vspace*{-1.05in}\hspace*{2.1in}
\begin{minipage}[r]{3.5in}
\footnotesize
{\bf FIGURE 1.} Gaugino/higgsino-loop diagrams giving rise to modifications 
to parameters in the potential. (Each dot indicates a helicity flip 
in the fermion propagator.)\\[.2in]
\hspace*{.2in} \rule{3in}{.02in} 
\normalsize
\end{minipage}

\section*{A new model with an exotic quark doublet}
Before going into discussion of a consistent 1-loop treatment, we first 
present a new model introduced in Ref.\cite{paper} that we believe
could be experimentally viable. Our new model has an extra pair of 
vectorlike quark doublets, $Q$ and $\bar{Q}$,
with the following couplings
\begin{equation}
{\cal L}_Q = M_Q  \bar{Q} Q + \lambda_Q \bar{t} \phi_1^{\dag} Q\; ,
\end{equation}
as an addition to the  two-Higgs-doublet SM  or MSSM. Note that 
$\phi_1^{\dag}$ is actually $H_d$, the Higgs (super)multiplet that gives
masses to the down-type quarks; and $\phi_2$ is  $H_u$. 
So, $T_3=-1/2$ component of $Q$, denoted by $T$,
has the same charge as the top quark and mixes with it after electroweak
(EW) symmetry breaking. The other part of the doublet is a
quark of electric $5/3$.
The 1-loop diagram, now with the gaugino/higgsino propagators
replaced by that of the quarks, leads to 
$\Delta\lambda_5(\sim 3\lambda_Q^2\lambda_t^2/16\pi^2)$ 
and could be very substantial for large Yukawa couplings. 
(Note that $\phi_1$ and $\phi_2$ vertices now have 
$\lambda_Q$ and $\lambda_t$ couplings respectively). 

The mass matrix of the $t$-$T$ system is given by
\begin{equation}
{\cal M}_t = \left( \begin{array}{cc} \lambda_t \lla \phi_2 \rra & 
\lambda_Q \lla \phi_1^{\dag} \rra \\  0 &   M_Q \end{array}\right)\; .
\end{equation}
Notice that the model actually has an effective KM phase to account for
weak CP phenomenology. Assuming $M_Q$ to be roughly around
the same order as the EW scale, the model can also easily get around
the $``$small $m_A$"  objection. Moreover, its modification to top quark
phenomenology would  be very interesting, and will provide an experimental
check on its viability. The $Q$-$\bar{Q}$ exotic quarks could naturally arise,
for example, as the only extra quarks from some interesting 
models with a SM-like chiral fermion spectra  embedding the three SM 
families in a intriguing way\cite{kg}.

\section*{A consistent 1-loop treatment and the Georgi-Pais theorem}
Here we give a brief discussion of our consistent 1-loop treatment for
both MSSM and the new model, particularly in relation to the
Georgi-Pais theorem\cite{gp}. Readers are referred to our original
paper\cite{paper} for more details.

The Georgi-Pais theorem states that radiative CP violation can occur if and 
only if there exists spinless bosons which are massless in the tree approximation.
Naively looking, the above radiative CP violation pictures violate the
theorem. There are some confusing statement in the literature in relation to
the situation. However, the theorem has a presumption, that no fine tuning
be allowed. Results from our consistent treatment of the two models discussed
to be sketched below show exactly that some form of fine tuning is
unavoidable for radiative CP violation to occur in both cases. There has also
been statements about the smallness of $m_{\scriptscriptstyle A}$ being
a necessary consequence of the Georgi-Pais theorem. Comparison between
the two models dicsussed here clearly illustrated that  is not true. Smallness
of $m_{\scriptscriptstyle A}$ in the MSSM cases is rather the  result of the 
small gauge couplings used to produce $\Delta\lambda_5$. 

\addtocounter{figure}{1}

\begin{figure}[t]
\begin{center}
\rule{5in}{.02in}
\end{center}
\vspace{1.2in}
\includegraphics{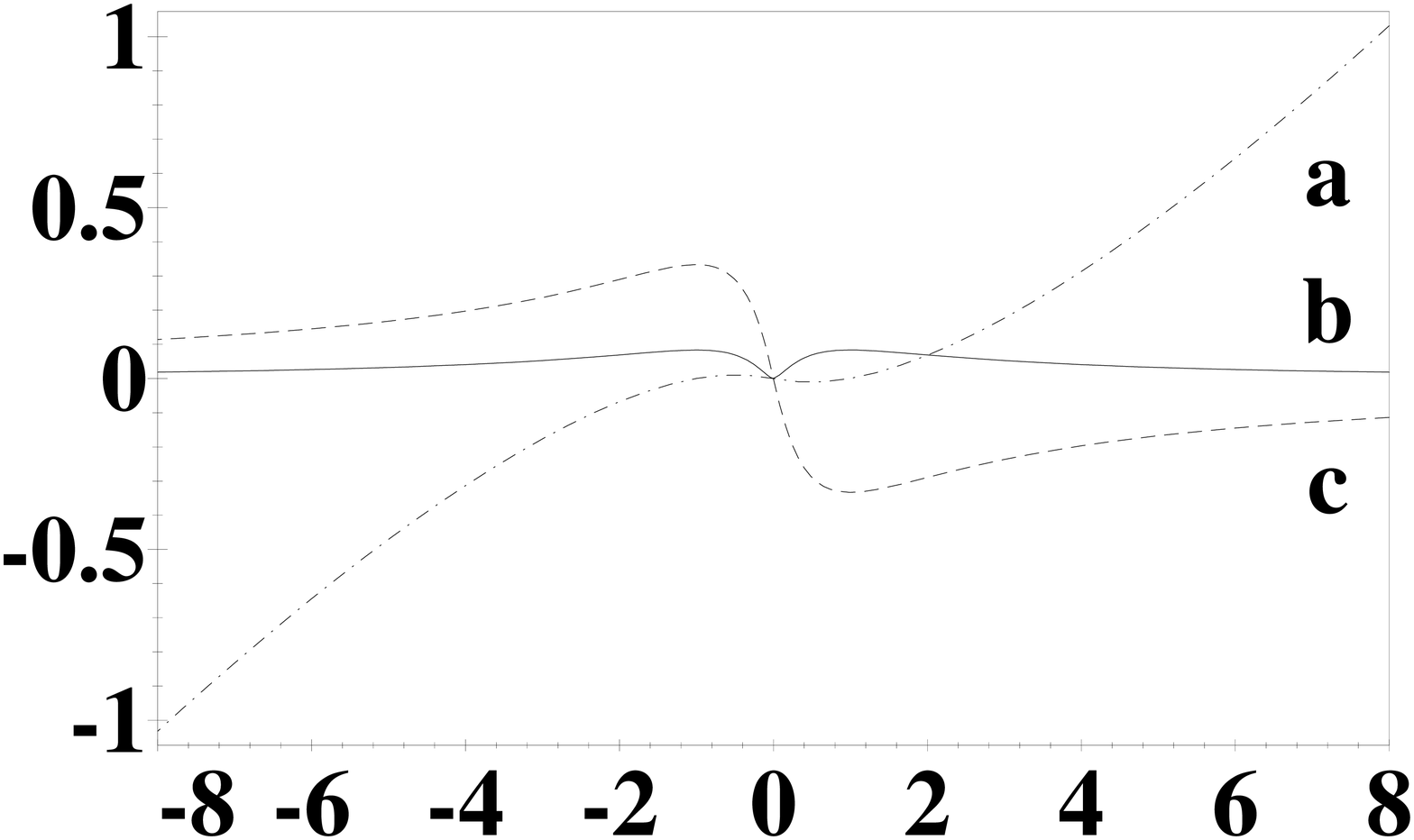}
\vspace{2in}
\caption{ Plots of radiative correction from chargino loop 
verses $m(=M_{\tilde{g}}/\mu)$, with mass mixing from EW symmetry neglected. 
(a)$\Delta m_3^2$($\overline{MS}$) in $25\times g^2\mu^2/16\pi^2$; 
(b)$\Delta \lambda_5$; (c)$\Delta \lambda_6$ ($=\Delta \lambda_7$);
both of the latter curves with values in $g^4/16\pi^2$. }
\vspace{.2in}
\begin{center}
\rule{5in}{.02in}
\end{center}
\end{figure}
\vspace{.2in}

A positive $\Delta\lambda_5$ is a necessary but {\it not sufficient} condition 
for radiative CP violation. This is clearly illustrated in our discussion of the 
scalar potential and the problem of the Maekawa picture.
In our opinion, it is at least of theoretical interest to see  
what the 1-loop gaugino/higgsino effect alone could do to the vacuum
solution of a supersymmetric 2HDM. A consistent treatment of
the 1-loop effect should of course take into consideration 
contributions to all the 10 parameters in the potential $V$. Recall that
another essential condition for the existence of the CP
violating vacuum solution is $|\Omega|<1$. 

We plot the numerical results of major interest in Fig.2.
The plots are for the chargino contributions only, as functions of
$m=M_{\tilde{g}}/\mu$, the gaugino-higgsino mass ratio.
Our results here presented give the 1-loop effect before EW
symmetry breaking, {\it i.e.} mass mixing between the gaugino and 
higgsino were not considered.  Further modifications due to the
symmetry breaking are not expected to change the general features. Here, the
neutralino contributions can simply be inferred from symmetry.

Taking the renormalized value of  $m_3^2$ as  a free parameter,
it has a magnitude that increases fast with
that of $m$ for $|m|>1$.  Obviously, some fine tuning is needed to get
$|\Omega|<1$, though a small window on $m$ always exists not too far
from $|m|=1$, for each sign of $\mu$, for a not too small  $\mu$. 
With EW-scale  $\mu$, 
$\Delta m/m$ of the admissible regions are of order $10^{-2}$, though
the severe fine tuning can be tamed by having small $\mu$.

\begin{figure}[b]
\begin{center}
\rule{5in}{.02in}
\end{center}
\vspace{1.5in}
\includegraphics{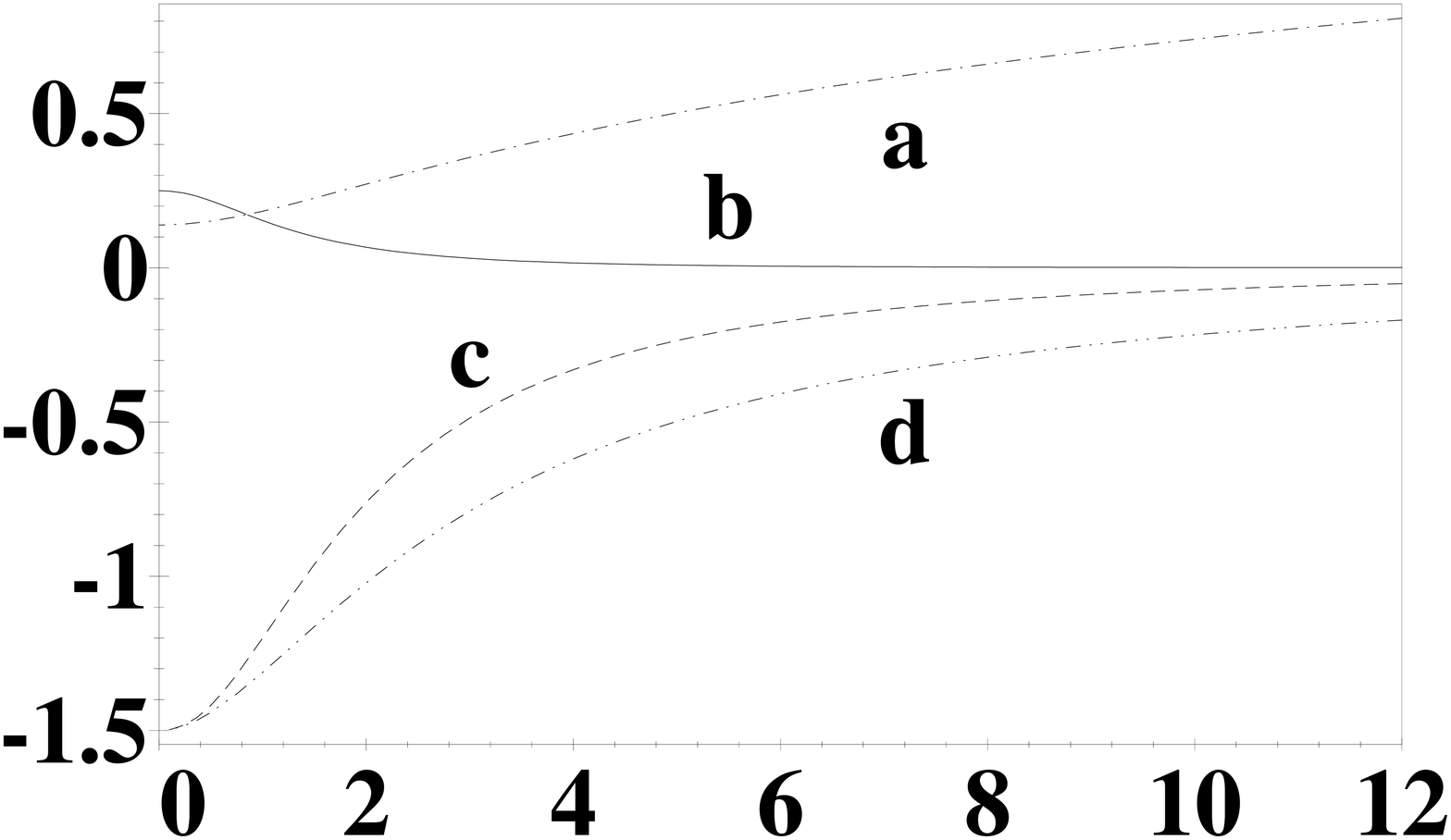}
\vspace{1.9in}
\caption{Plots of radiative correction from $t$-$T$ loop
verses $m(=M_Q/v_2)$.  (a)$\Delta m_3^2$($\overline{MS}$) 
in $10\times 3\lambda_t^2v_2^2/16\pi^2$; (b)$\Delta \lambda_5$; 
(c)$\Delta \lambda_6$; (d)$\Delta \lambda_7$; all of the latter three curves with values in $3\lambda_t^4/16\pi^2$; ($\lambda_t=\lambda_Q$, $\tan{\beta}=1$ assumed).} 
\vspace{.2in}
\begin{center}
\rule{5in}{.02in}
\end{center}
\end{figure}
\vspace{.2in}

In the our new model, though the new quark doublet
$Q$ has mass before EW symmetry breaking, a similar set of
1-loop diagrams, as those given in Fig.1, can only be completed when the
EW breaking mass of the top and its mixing with $T$ are taken into 
consideration. But then the plausibly large Yukawa couplings
give rise to substantial results.  In Fig.3, we presented some numerical
results. The plots use again $m$ as parameter which in this case
denotes $M_Q/v_2$. For simplicity, we assume $\lambda_Q = \lambda_t$.

Here , $\Delta m_3^2$
($\overline{MS}$) always has the opposite sign to that of
$\Delta\lambda_6$ or $\Delta\lambda_7$, making a naive use of the
value to fit in the $|\Omega|<1$ condition impossible.
Hence, for SCPV to occur, a tree level  $m_3^2$ value is needed. 
While the  $m^2_{3(tree)}$ then has to
be chosen to roughly match the $t$-$T$ 1-loop effect, the large
Yukawa couplings make this relatively natural, as a value of the order 
$v_2$ is all required. Accepting that,   $\Delta m^2_{3(tree)}/m^2_{3(tree)}$ 
of the solution region is not quite small, 
$\mathrel{\raise.3ex\hbox{$>$\kern-.75em\lower1ex\hbox{$\sim$}}} .35$ 
for $m\leq 1$, representing only a moderate fine tuning. 
The small $m_3^2$ required is natural, as its zero limit provides
the scalar potantial with an extra Peccei-Quinn type symmetry.

\section*{conclusion}
We have performed a consistent 1-loop analysis of the feasibility of radiatively
induced SCPV, for both the MSSM and our proposed new model with a pair of
vector-like exotic quark doublets.  Our results clarified some issues concerned.
The new model, especially a SUSY version, is expected to be 
phenomenologically interesting. Though a complete studies of its various
features and experimental viability still have to be performed, we have
illustrated at least how it can easily overcome the objections to the 
corresponding scenario within MSSM. Further studies of the model is
under progress.

\end{document}